\documentstyle[12pt]{article}
\author{Stefan ~G\"ullenstern \\
Max-Planck-Institut f\"ur Kernphysik, Postfach 103 980 \\
D-69029 Heidelberg, Germany
\and Pawe\l\ G\'ornicki \\
Institute of Physics, Polish Academy of Sciences\\
and\\
Center for Theoretical Physics, Polish Academy of Sciences\\
Al. Lotnik\'ow 32/46,
PL-02-668 Warsaw, Poland
\and Lech Mankiewicz \\
Nicolaus Copernicus Astronomical Center, ul. Bartycka 18 \\
PL-00-716 Warsaw, Poland
\and Andreas Sch\"afer\\
Institut f\"ur Theoretische Physik, Universit\"at Frankfurt\\
D-60054 Frankfurt am Main, Germany}
\title{{\sc Sphinx}: Monte Carlo Program for Polarised Nucleon-Nucleon
Collisions}
\begin{document}
\maketitle

\noindent
{\LARGE PROGRAM SUMMARY}

\vskip 0.5cm
\begin{trivlist}
\item[\it Title of program:] {\sc Sphinx} -- HEP MONTE CARLO
\item[\it Catalogue number:]
\item[\it Program available from:]
\item[\it Computer:] DEC Station 5000/260; other machines
with {\tt FORTRAN~77} and sufficient capacity should be able to run
this  program
\item[\it Installation:] Institut for Theoretical Physics, University
of Frankfurt, Frankfurt/Main
\item[\it Operating system:] tested under {\tt UNIX} but does not depend
on the particular operating system
\item[\it Programing language used:] {\tt FORTRAN~77}
\item[\it High speed storage required:] 5 Mbytes
\item[\it Number of bits per byte:] 8
\item[\it Number of lines in combined test deck:] 19296
\item[\it CPC subprograms used:] {\sc Jetset~7.4}
\item[\it Keywords:] polarized nucleon-nucleon scattering, high
energy physics, Monte Carlo simulation
\item[\it Nature of physical problem:]
This program can be used to simulate polarised nucleon - nucleon
collisions at high energies. Spins of colliding particles are
taken into account. The program allows to calculate cross
sections for various processes.
\item[\it Method of solution:]
The existing Monte Carlo program {\sc Pythia~5.7} has been modified to
incorporate spin effects. The program incorporates all features
of {\sc Pythia}.
\item[\it Restrictions on the complexity of the problem:]
Spins of colliding nucleons must be parallel to the collision
axis.
\item[\it Typical running time:] $\approx 0.1$ sec CPU per event on
DEC Station 5000/260
\item[\it References:]
Reader has to refer to {\sc Pythia} manual \cite{pyman} for
additional details.
Further informations can be obtained from: Prof.\ A.\
Sch\"afer, Institut f\"ur Theoretische Physik, Universit\"at Frankfurt
a.\ M., Robert Mayer-Strasse 10, D-60054 Frankfurt am Main, Germany
\end{trivlist}
\pagebreak

\noindent
{\LARGE LONG WRITE-UP}

\section{Introduction\label{introduction}}
The detailed investigation of hadronic spin effects, most notably in
deep-inelastic lepton-nucleon scattering, has proven to allow for extremly
sensitive tests of QCD. One of the most interesting issues is the possible
contribution of the axial-vector anomaly. This anomaly plays a central role
in modern quantum-field-theory and spin effects might offer the only
possibility to actually detect it. To estimate its contribution one has to
know the distribution function for polarised gluons $\Delta g(x,Q^2)$. It is
argued that the total spin carried by gluons could be large, because
$\alpha_s(Q^2)\Delta g(Q^2)$, with
$\Delta g(Q^2)=\int_0^1{\rm d}x\ \Delta g(x,Q^2)$,
is a renormalisation group invariant, such that if $\Delta g(Q^2)$ is a few
percent at the constituent quark level it could be very sizeable at
large $Q^2$. In deep-inelastic scattering it is impossible to distinguish
between the regular quark contribution and the gluon contribution via the
anomaly. There is actually no conceptually sound way to separate
them. In exclusive or semi-inclusive lepton-hadron scattering
channels it is in principle possible to discern a gluonic and a quark
contribution, but the by far most direct and conceptually clean way are
polarised proton-proton collisions. RHIC \cite{rhic} offers the
option to study
polarised proton collisions up to $\sqrt{s}=300\ {\rm GeV}$ which
would be ideal for this purpose. To analyse this option in detail, to
plan the experiments, and to determine the reachable precission a
detailed Monte Carlo program is needed, which we provide with this
publication. This program can handel, however, only longitudinal
polarisation. All polarisation effects in proton-proton reactions share
the same problem. As only a small fraction of the partons is polarised
the signal to background ratios are typically of the order percent,
such that a detailed simulation is necessary to extract quantitative results.

Because a state-of-art Monte
Carlo program is usually the result of many years of development
we modified an existing, widly used program
called {\sc Pyhtia} \cite{pyman}. We have been interested
only in those parts of the program that correspond
to nucleon-nucleon collisions. Consequently only these routines were changed.
The resulting polarised version of {\sc Pythia} is named {\sc Sphinx},
an acronym
for {\em {\bf S}imulator of {\bf P}olarised {\bf H}adronic {\bf
IN}teract({\bf X})ions}.

It is interesting to note that {\sc Sphinx} is able to function in unpolarised
mode as well. In that case it is functionally equivalent to {\sc Pythia}.
This means that results will be the same but because some data structures
inside the program have been modified the program requires more
space.

The main problem was to introduce the changes in a way compatible with
the basic structure of {\sc Pythia}, e.\ g.\ the way in which {\sc
Pythia} evaluates the cross sections or simulates the effects of
higher order QCD corrections.
{\sc Pythia} is based on the parton
model. The main assumption is that the nucleon may be represented
by an incoherent mixture of quarks and gluons carrying portions
of nucleon momentum. Their momentum distributions are described by parton
distributions.  In the parton model the nucleon is composed of quarks and
antiquarks of various
flavours and of gluons. In the polarised
case each parton comes in one of two helicities:
righthanded and lefthanded. Therefore the number of parton components
has to be doubled, corresponding to the two degrees of freedom for
longitudinal polarisation.
The most natural idea was
to generalize the notion of flavour to account for the polarisation.
So instead of d-quark we have lefthanded d-quark and righthanded d-quark
etc. Because in {\sc Pythia} gluons are treated as a special ``flavour''
the scheme
applies to them as well. As far as the general structure of the program
is concerned
the only complication of the polarised code is that we have more
``flavours'' and this is rather easy to implement.

Polarisation is followed in {\sc Sphinx} only up to the hard partonic
interaction, i.\ e.\ the hadronic cross section and the higher order
corrections in the initial state shower are evaluated spin{\em dependently}
whereas all final state interactions as fragmentation and
decays are treated spin{\em independently}. No theoretical model for polarised
fragmentation exists. It is,
however, generally accepted that it should not depend on longitudinal
polarisation. For transverse polarisation the situation would,
however, be different. One knows experimentally that substantial
effects exist, e.\ g.\ a strong coupling between transverse spin and
transverse momentum.

The polarisation effects have been implemented in the
following parts of the program:
\begin{enumerate}
\item Parton distributions. At the moment we supply six sets
of polarised parton distributions: Altarelli-Stirling \cite{altsti},
two parametrisations from a Ross-Roberts article \cite{rosrob} and
three sets from Gehrmann\&Stirling
\cite{gehrmann}.
First order Altarelli-Parisi \cite{altpar} evolution is taken into
account in all cases.
\item Hard processes. The processes currently implemented in the
polarised mode are summarized in Table \ref{processes}.
Other processes may be used
as well but the results will be averaged over polarisations.
The cross sections for all polarised processes are given in the literature
\cite{crosssec},
and we checked them.
\item Initial state showering. The polarised case was not discussed in
this context
before. We obtained all necessary formulae.
\item Documentation. The event listing has been modified such that it
displays polarisation information about the interacting particles up
to the hard partonic interaction.
\end{enumerate}

\begin{table}
\label{processes}
\caption{List of processes implemented in the polarised mode.}
\begin{center}
\begin{tabular}{|r|l|p{8cm}|}
\hline
{\tt ISUB} & Process & Comment \\
\hline\hline
1 & $q_i\bar{q}_j\rightarrow \gamma^* / Z^0 $ & quark-antiquark
annihilation into virtual $ \gamma^* / Z^0$ \\
\hline
2 & $q_i\bar{q}_j\rightarrow W^\pm $ & annihilation into charged
vector boson \\
\hline
11 & $q_iq_j\rightarrow q_iq_j$ &
(anti-)quark -- (anti-)quark scattering; annihilation diagram not
included \\
\hline
12 & $q_i\bar{q}_i\rightarrow q_k\bar{q}_k$ &
annihilation process \\
\hline
13 & $q_i\bar{q}_i\rightarrow g g$ &
annihilation into gluon pair \\
\hline
14 & $q_i\bar{q}_i\rightarrow g \gamma$ &
annihilation into gluon and prompt $\gamma$ \\
\hline
15 & $q_i\bar{q}_i\rightarrow g Z^0$ & annihilation into gluon
and $Z^0$ \\
\hline
16 & $q_i\bar{q}_i\rightarrow g W^\pm$ & annihilation into
gluon and $W^\pm$ \\
\hline
18 & $q_i\bar{q}_i\rightarrow \gamma \gamma$ &
annihilation into $\gamma$ pair \\
\hline
19 & $q_i\bar{q}_i\rightarrow \gamma Z^0$ & annihilation into
$\gamma$ and $Z^0$ \\
\hline
20 & $q_i\bar{q}_i\rightarrow \gamma W^\pm$ & annihilation into
$\gamma$ and $W^\pm$ \\
\hline
28 & $q_ig\rightarrow q_ig$ &
(anti-)quark -- gluon scattering \\
\hline
29 & $q_ig\rightarrow q_i\gamma$ &
prompt $\gamma$ production in (anti-)quark -- gluon scattering \\
\hline
30 & $ q_ig \rightarrow q_iZ^0$ & $Z^0$ production in
(anti-)quark -- gluon scattering \\
\hline
31 & $ q_ig \rightarrow q_jW^\pm$ & $W^\pm$ production in
(anti-)quark -- gluon scattering \\
\hline
53 & $g g\rightarrow q_k\bar{q}_k$ &
gluon fusion \\
\hline
68 & $g g\rightarrow g g$ &
gluon -- gluon scattering \\
\hline \hline
\end{tabular}
\end{center}
\end{table}

Hereafter we do not try to explain the structure of {\sc Pythia~5.7}
and we indicate
only the most significant modifications which lead to {\sc Sphinx}.
Readers unfamiliar with {\sc Pythia}
should first consult the appropriate manual \cite{pyman}.
\section{From {\sc Pythia} to {\sc Sphinx}: modifications\label{main}}
Our program is an extension of {\sc Pythia} version 5.7.
The {\sc Pythia~5.7} program heavily relies on {\sc Jetset~7.4}
subroutines \cite{jetset}.
These subroutines are mainly used to simulate processes that take
place after the hard partonic interaction, i.e. fragmentation and
decays, or provide spinindependent manipulations as for example
Lorentz-transformations.
Because polarisation
effects in the final state are neglected in {\sc Sphinx},
the {\sc Jetset} program could be left untouched. Only the set-up of an event
listing which is also a task of {\sc Jetset} in a {\sc Pythia} run has
been taken over from {\sc Sphinx} in the polarised mode. The
event listing in {\sc Sphinx} provides then in addition information
about the polarisation flow.
For
that purpose two new subroutines have been added which are
modifications of the corresponding {\sc Jetset} routines (see below).
By this means {\sc Sphinx} provides a suitable interface to {\sc
Jetset} such that the programs
may be linked without any modifications in the latter.
There is only one exception: the {\sc Jetset} subroutine {\tt LUGIVE},
which returns the values of all {\sc Jetset} {\em and} {\sc Pythia}
common block variables cannot be longer used, because in {\sc Sphinx}
the dimensions of some {\sc Pythia} arrays have been enlarged so that
the formats no longer match.

In many places inside {\sc Pythia} partonic data are stored in arrays indexed
by flavour. Inclusion of helicities means doubling the number of ``flavours''
as described in Section \ref{introduction}.
For that reason we had to enlarge the arrays.
The unpolarised mode has been preserved throughout the program
and works exactly as the original
code.
All modifications are implemented in a treelike structure with several
branching points where the user may choose between the
polarised and unpolarised modes. The default is always unpolarised.
An interesting point is that one can combine polarised and unpolarised
treatments for different processes and for various stages  of the same
process. This has to be done with care and one has to be aware of this
not always being justified from the
physical point of view.
The branching points were
implemented by means of the {\tt IF-ELSE} structure
which directs the program flow according to a specific switch.
Each functional part (usually a subroutine) has its own {\em local}
variable
called {\tt IPOL}. This variable controls the mode and its value depends
on some
parameters\footnote{See Table \ref{parameters}} {\tt MSTP(x)} and {\tt
NSUB(x)}
to be defined by the user in the main program.
For example, it is
possible to run the code with polarised partonic cross
section and polarised parton distributions but unpolarised initial state
showering etc. This solution provides flexibility but has to be
used carefully.
The major disadvantage of {\tt IF-ELSE} constructions is that parts of
the code are multiplied. This makes servicing
of the code more cumbersome because one has to introduce the same
change in different places at once. In addition the resulting code
became quite lengthy.

{\sc Sphinx} is as {\sc Pythia} a ``slave-system'', i.\ e.\ it
consists only of callable subroutines where two of them ({\tt
PYINIT} and {\tt PYEVNT}) have to be called by the user to
perform the event generation and a few other, e.\ g.\ {\tt DPLIST},
{\tt PYSTAT}, etc., could be called to obtain further event information.
The rest of the subroutines is of internal use. Therefore the user has
to supply
a main program where all relevant parameters and switches have to be
specified and these subroutines have to be called. Because this
structure is the same as in {\sc Pythia} the reader is referred to
\cite{pyman} again for details about the general {\sc Pythia}
parameters. In this article only the new parameters in {\sc Sphinx}
are discussed.
The modifications made in the individual subroutines are described in Section
\ref{subroutines}.
In Table \ref{parameters} we list all new parameters introduced
to control the polarised mode.
Finally in Section \ref{example} examples for a main program
and the result of the corresponding test runs are presented.
\begin{table}
\label{parameters}
\caption{Parameters controlling the polarised mode}
\begin{center}
\begin{tabular}{|l|rp{8cm}|c|c}
\hline
Parameter &&Description & Default \\
\hline\hline
{\tt MSTP(171)} && beam polarisation & 0\\
&=0:& unpolarised  & \\
&=1:& polarisation in $+z$ direction &\\
&=2:& polarisation in $-z$ direction &\\
\hline
{\tt MSTP(172)} && target polarisation & 0\\
&=0:& unpolarised  & \\
&=1:& polarisation in $+z$ direction &\\
&=2:& polarisation in $-z$ direction &\\
\hline
{\tt MSTP(175)} && use of polarised parton distribution in polarised
initial state shower ({\tt MSTP(176)=1})& 1\\
&=0:& unpolarised distribution; for testing only, {\em do not use!}&\\
&=1:& polarised distribution&\\
\hline
{\tt MSTP(176)} && initial state showering mode & 0\\
&=0:& unpolarised & \\
&=1:& polarised & \\
\hline
{\tt MSTP(177)} && set of polarised parton distributions
$\Delta q$ and $\Delta g$ used; in addition one has
to specify unpolarised set as in standard {\sc Pythia} & 0\\
&=0:& $\Delta q =0$ and $\Delta g=0$ (no polarisation) & \\
&=1:& fake polarisation, built up from unpolarised distribution
according to $\Delta q = \frac{\tt MSTP(178)}{100}q$ & \\
&=2:& Altarelli-Stirling parametrization \cite{altsti}; data
file {\tt altsti.dat} required & \\
&=3:& Ross-Roberts parametrization \cite{rosrob} set d & \\
&=4:& Ross-Roberts parametrization \cite{rosrob} set a; data
file {\tt rosroa.dat} required & \\
&=5:& Gehrmann-Stirling parametrization \cite{gehrmann} set a; data
file {\tt partons.dat} required & \\
&=6:& Gehrmann-Stirling parametrization \cite{gehrmann} set b; data
file {\tt partons.dat} required & \\
&=7:& Gehrmann-Stirling parametrization \cite{gehrmann} set c; data
file {\tt partons.dat} required & \\
&=8:& fake polarisation, for testing only, {\em do not use!}& \\
\hline
\end{tabular}
\end{center}
\end{table}
\setcounter{table}{1}
\begin{table}
\caption{Parameters controlling the polarised mode, continued}
\begin{center}
\begin{tabular}{|l|rp{8cm}|c|c}
\hline
Parameter &&Description & Default \\
\hline\hline
{\tt MSTP(178)}&& percentage of fake polarisation for {\tt MSTP(177)=1} & 0\\
\hline
{\tt MSTP(180)} && mode selection (master switch) & 0\\
&=0:& unpolarised mode; this value overrides all other
polarisation switches & \\
&=1:& polarised mode  & \\
\hline
{\tt NSUB(ISUB)} && mode for subprocess ISUB & 0 \\
&=0:& unpolarised treatment &\\
&=1:& polarised treatment &\\
\hline
{\tt NSEL} && menue of polarised processes & 0 \\
&=1:&  {\tt ISUB = 11,12,13,28,53,68} switched on&\\
&=10:& {\tt ISUB = 14,18,29} switched on&\\
&=11:& {\tt ISUB = 1} switched on&\\
&=12:& {\tt ISUB = 2} switched on &\\
&=13:& {\tt ISUB = 15,30} switched on &\\
&=14:& {\tt ISUB = 16,31} switched on&\\
\hline
\end{tabular}
\end{center}
\end{table}
\begin{table}
\label{internal}
\caption{Internal variables storing polarisation information}
\begin{center}
\begin{tabular}{|l|rp{8cm}|c|c}
\hline
Variable &&Description & Com.\ Block \\
\hline\hline
{\tt MINT(311)} && beam helicity & {\tt PYINT1}\\
&=0:& unpolarised  & \\
&=1:& positive helicity &\\
&=2:& negative helicity&\\
\hline
{\tt MINT(312)} && target helicity & {\tt PYINT1}\\
&=0:& unpolarised  & \\
&=1:& positive helicity &\\
&=2:& negative helicity&\\
\hline
{\tt MINT(313)} && helicity of shower initiator on beam side& {\tt PYINT1}\\
&=0:& unpolarised  & \\
&=1:& positive helicity &\\
&=2:& negative helicity&\\
\hline
{\tt MINT(314)} && helicity of shower initiator on target side& {\tt PYINT1}\\
&=0:& unpolarised  & \\
&=1:& positive helicity &\\
&=2:& negative helicity&\\
\hline
{\tt MINT(315)} && helicity of hard interacting parton on beam side&
{\tt PYINT1}\\
&=0:& unpolarised  & \\
&=1:& positive helicity &\\
&=2:& negative helicity&\\
\hline
{\tt MINT(316)} && helicity of hard interacting parton on target side&
{\tt PYINT1}\\
&=0:& unpolarised  & \\
&=1:& positive helicity &\\
&=2:& negative helicity&\\
\hline
{\tt MSTP(179)}&& switch off polarisation temporarely in {\tt PYSIGH}
and {\tt PYSTFU} resp. & {\tt PYPARS}\\
&=0:& no action   & \\
&=1:& switch off polarisation &\\
\hline
\end{tabular}
\end{center}
\end{table}
\setcounter{table}{2}
\begin{table}
\caption{Internal variables storing polarisation information, continued}
\begin{center}
\begin{tabular}{|l|rp{7cm}|c|c}
\hline
Variable &&Description & Com.\ Block \\
\hline\hline
{\tt ISIGH(1000,6)} && hard scattering information of {\tt I}th line & {\tt
PYINT3}\\
{\tt ISIGH(I,1)}&& particle code of {\tt I}th line on beam side  & \\
{\tt ISIGH(I,2)}&& particle code of {\tt I}th line on target side  & \\
{\tt ISIGH(I,3)}&& colour flow  & \\
{\tt ISIGH(I,4)}&& helicity of {\tt I}th line on beam side  & \\
{\tt ISIGH(I,5)}&& helicity of {\tt I}th line on target side  & \\
{\tt ISIGH(I,6)}&& not used  & \\
\hline
{\tt KD(I)} && polarisation/helicity of {\tt I}th line & {\tt DPYPOL}\\
&=0:& no polarisation/helicity   & \\
&=1:& positive polarisation/helicity   & \\
&=2:& negative polarisation/helicity   & \\
\hline
{\tt XSFX(2,-40:40,0:2)} && $x$ times parton distribution for given $x$
and $Q^2$ of flavour {\tt KFL = -40:40} and helicity {\tt KFLD = 0:2} on beam
side ({\tt JT=1}) and target side
({\tt JT=2}) resp.& {\tt PYINT3}\\
{\tt XSFX(JT,KFL,0)}&& unpolarised   & \\
{\tt XSFX(JT,KFL,1)}&& positive helicity   & \\
{\tt XSFX(JT,KFL,2)}&& negative helicity   & \\
\hline
\end{tabular}
\end{center}
\end{table}

\section{Common Blocks and Subroutines\label{subroutines}}
In the following the modified and the new
subroutines that have been specifically created for {\sc Sphinx} are
described in more detail. The general tasks of the subroutines
themselves as well as the unchanged parameters and variables are not
explained, because they are the same as in {\sc Pythia}. The reader is
asked again if occasion arises to consult \cite{pyman} to obtain the
needed information. We restrict our explanation to the new aspects in
{\sc Sphinx}.
The purpose of the modifications is
indicated and the new parameters, switches, and internal variables are
listed. Meaning and possible values of the new parameters are
given in Table \ref{parameters}.
To incorporate polarisation the following common blocks have been
enlarged and replace the
corresponding {\sc Pythia} common blocks or are added:
\begin{itemize}
\item {\tt COMMON/PYINT3/XSFX(2,-40:40,0:2),ISIG(1000,6),SIGH(1000)}
\item {\tt
COMMON/PYSUBS/MSEL,NSEL,MSUB(200),NSUB(200),KFIN(2,-40:40),
CKIN(200)}
\item {\tt COMMON/DPYPOL/KD(4000)}
\end{itemize}
Information about the new
internal variables and enlarged arrays can be found in Table \ref{internal}.
In addition
it is shown how the local polarisation switch {\tt IPOL} is built up
in the different subroutines. Only the
polarised case ({\tt IPOL=1}) will be discussed, because in the
unpolarised case ({\tt IPOL=0}) each subroutine works exactly as the
corresponding {\sc Pythia} subroutine.
The not mentioned subroutines of {\sc Sphinx} are the same as in {\sc Pythia}.

\paragraph{\tt MAIN PROGRAM}\hfill\break
\noindent{\bf Purpose:} to set up the polarised event generation. The
variables
which have to been set are listed in Table \ref{parameters}.\hfill\break
{\bf Remarks:} Examples of a main program are given in Section
\ref{example}.

\paragraph{\tt SUBROUTINE PYINIT(FRAME,BEAM,TARGET,WIN)}\hfill\break
\noindent{\bf Purpose:} to display {\sc Sphinx} header; to check
partially the availability of the
desired polarisation
scenario, i.\ e.\ to control that the master switch for polarisation
{\tt MSTP(180)} is set properly, the selected partonic
subprocesses can be treated polarised and to control and compose the
polarisation
menue via {\tt NSEL}; to call {\tt DPLIST} instead of {\tt
LULIST} (see below).\hfill\break
{\bf New parameters:} {\tt MSTP(180)}, {\tt NSEL}, {\tt
NSUB(ISUB)}\hfill\break
{\bf Internal Polarisation switch:} {\tt IPOL=MSTP(180)}\hfill\break
{\bf Remarks:} If a not allowed scenario has been chosen, the
programs stops with an appropriate error message.

\paragraph{\tt SUBROUTINE PYEVNT}\hfill\break
\noindent{\bf Purpose:} to start polarised event generation;
to call {\tt DPEDIT} instead of {\tt LUEDIT} (see below).\hfill\break
{\bf New parameters:} {\tt MSTP(180)}\hfill\break
{\bf Internal polarisation switch:} {\tt IPOL=MSTP(180)}

\paragraph{\tt SUBROUTINE PYINKI(CHFRAM,CHBEAM,CHTARG,WIN)}\hfill\break
\noindent{\bf Purpose:} to check availability of the desired hadronic
polarisation
scenario, i.\ e.\ to control that the selected hadron
can be treated polarised and to verify that the polarisation
is longitudinal; to store the polarisation of beam and target for the
event listing in {\tt KD(1)} and {\tt KD(2)} and for internal use in
{\tt MINT(311)} and {\tt MINT(312)}.\hfill\break
{\bf New parameters:} {\tt KD(I)}, {\tt MSTP(171)}, {\tt MSTP(172)},
{\tt MSTP(180)} \hfill\break
{\bf New internal variables:} {\tt MINT(311)}, {\tt MINT(312)}\hfill\break
{\bf Internal polarisation switch:} {\tt IPOL=MSTP(180)}\hfill\break
{\bf Remarks:} At present only nucleons and hyperons and their
antiparticles can be treated polarised.
If a not allowed scenario has been chosen, the
programs stops with an appropriate error message.

\paragraph{\tt SUBROUTINE PYRAND}\hfill\break
\noindent{\bf Purpose:} to adapt {\tt PYRAND} to the new
environment -- all relevant arrays which have been enlarged or added to the
common blocks in other subroutines are modified here as well; to
extend event shape selection to incorporate helicities;
to store
helicities of the partons entering the hard interaction according to
\begin{itemize}
\item {\tt MINT(313)}: helicity of the beam parton for use in the initial
state showering subroutine;
\item {\tt MINT(314)}: helicity of the target parton for use in the initial
state showering subroutine;
\item {\tt MINT(315)}: helicity of the beam parton;
\item {\tt MINT(316)}: helicity of the target parton.
\end{itemize}
\hfill\break
{\bf New parameters:} {\tt MSTP(180)}, {\tt NSUB(ISUB)} \hfill\break
{\bf New internal variables:} {\tt MINT(313)}, {\tt MINT(314)}, {\tt
MINT(315)}, {\tt MINT(316)}\hfill\break
{\bf Internal polarisation switch:} {\tt IPOL=MSTP(180)$\times$NSUB(ISUB)}
\hfill\break
{\bf Remarks:}
Note that {\tt MINT(313)=MINT(315)} and {\tt MINT(314)=MINT(316)} but the
values of {\tt MINT(313)} and {\tt MINT(314)} are changed later by
the initial state shower in {\tt PYSSPA}.

\paragraph{\tt SUBROUTINE PYSCAT}\hfill\break
\noindent{\bf Purpose:} to adopt {\tt PYSCAT} to the new environment
(see {\tt PYRAND}); to store helicities of the partons entering the
hard interaction; to fill lines 1, 2 and 5, 6 in the event listing
with polarisation information (see below).
\hfill\break
{\bf New parameters:} {\tt KD(I)}, {\tt MSTP(180)}, {\tt NSUB(ISUB)}
\hfill\break
{\bf New internal variables:} {\tt MINT(315)}, {\tt MINT(316)}\hfill\break
{\bf Internal polarisation switch:} {\tt
IPOL=MSTP(180)$\times$NSUB(ISUB)}

\paragraph{\tt SUBROUTINE PYSSPA(IPU1,IPU2)}\hfill\break
\noindent{\bf Purpose:} to perform polarised initial state
showering, helicity dependent GLAP evolution equations are used in the
backward evolution algorithmus; to enlarge all relevant array in an
appropriate manner to incorporate polarisation;
to check proper selection of the
polarised initial state shower
scenario, i.\ e.\ to control that the {\tt MSTP(175)} and {\tt
MSTP(176)} are set correctly; to store the helicities of initial state
shower initiators ({\tt MINT(313)}, {\tt MINT(314)}).\hfill\break
{\bf New parameters:} {\tt KD(I)}, {\tt MSTP(171)}, {\tt MSTP(172)},
{\tt MSTP(175}, {\tt MSTP(176)}, {\tt MSTP(180)}, {\tt NSUB(ISUB)}
\hfill\break
{\bf New internal variables:} {\tt MINT(313)}, {\tt MINT(314)}, {\tt
MINT(315)}, {\tt MINT(316)}\hfill\break
{\bf Internal polarisation switch:}
{\tt IPOL=MSTP(180)$\times$NSUB(ISUB)$\times$MSTP(176)}
{\bf Remarks:} At the present stage only QCD shower can be treated
polarised, QED showering has to be done in the unpolarised manner.
The combination {\tt MSTP(175)=0} and {\tt MSTP(176)=1} allows to
simulate {\em polarised} showering with the use of {\em unpolarised}
parton distributions. This option is just for testing and should not
be selected by the user!
If {\tt MSTP(175)} or {\tt MSTP(176)} are set improperly, the
programs stops with an appropriate error message. The internal
variables
{\tt MINT(313)} and {\tt MINT(314)} are changed to their final values
in this subroutine.

\paragraph{\tt SUBROUTINE PYMULT(MMUL)}\hfill\break
\noindent{\bf Purpose:} to switch off polarisation in {\tt PYSIGH}
(set {\tt MSTP(179)=1} temporally)
when called from {\tt PYMULT} even in a polarised run, because
multiple interaction cannot be treated polarised at the moment.
\hfill\break
{\bf New parameters:} {\tt MSTP(179)}

\paragraph{\tt SUBROUTINE PYREMN(IPU1,IPU2)}\hfill\break
\noindent{\bf Purpose:} to adopt {\tt PYREMN} to the new environment
(see {\tt PYRAND}); to fill lines 3, 4 in the event listing
with polarisation information (see below).
\hfill\break
{\bf New parameters:} {\tt KD(I)}

\paragraph{\tt SUBROUTINE PYSIGH}\hfill\break
\noindent{\bf Purpose:} to evaluate the helicity dependent hadronic
cross sections by convolution of the helicity dependent parton
distributions with the helicity dependent partonic cross sections; to
supply the subroutine with the helicity dependent partonic cross
sections.\hfill\break
{\bf New parameters:} {\tt MSTP(171)}, {\tt MSTP(172)}, {\tt
MSTP(179)}, {\tt MSTP(180)}, \hfill\break{\tt
NSUB(ISUB)}\hfill\break
{\bf Internal Polarisation switch:} \hfill\break{\tt
IPOL=MSTP(180)$\times$NSUB(ISUB)$\times$(1-MSTP(179))}\hfill\break
{\bf Remarks:} {\tt PYSIGH} will always run in the unpolarised mode
when it is called by {\tt PYMULT} which sets {\tt MSTP(179)=1} in {\tt
PYSIGH} temporally. Evaluating the spindependent hadronic cross
sections one has to notice that the hadrons are specified according to
the spin, whereas the partons are labelled by their helicities.
The spin is defined relative to the
collision axis and the beam is assumed to move in the positive
direction. For that reason the helicities at the target side are
opposite to the polarisations. Hence the target labels are reversed in
the convolution in comparison to the beam labels.
The parton distributions are passed from {\tt PYSTFU} to {\tt PYSIGH}
through the array
{\tt XPQ(KFL,KFLD)} (see below) and stored in the array
{\tt XSFX(N,KFL,KFLD)}, where {\tt KFLD} denotes the helicity.
{\tt ISIG(N,I)} contains the information about the Nth-line in the
event listing. The new entries {\tt ISIG(N,4)} and {\tt ISIG(N,5)}
store the helicities of the partons at the beam and target side
respectively. {\tt ISIG(N,6)} is reserved but not used at the moment.

\paragraph{\tt SUBROUTINE PYSTFU(KF,X,Q2,XPQ)}\hfill\break
\noindent{\bf Purpose:} to evaluate the helicity dependent parton
distributions for given flavour ({\tt KF}), x ({\tt X}), and $Q^2$
({\tt Q2}) according to the selected parametrisations.\hfill\break
{\bf New parameters:} {\tt MSTP(177)}, {\tt MSTP(178)}, {\tt
MSTP(179)}, {\tt MSTP(180)}, \hfill\break{\tt
NSUB(ISUB)}\hfill\break
{\bf Internal Polarisation switch:} {\tt
IPOL=MSTP(180)$\times$(1-MSTP(179))}\hfill\break
{\bf Remarks:}
Call of {\tt PYSTFU} returns x times the parton distribution functions for
given flavour, $x$,  and $Q^2$ for both helicities and an averaged
(unpolarised) value.
The values are stored in the
array {\tt XPQ(KFL,KFLD)} which has been enlarged from {\tt XPQ(-25:25)} to
{\tt XPQ(-25:25,0:2)}. Row {\tt XPQ(KFL,0)}
contains the unpolarised distributions, {\tt XPQ(KFL,1)} distributions
corresponding to the positive
helicity (relative to the hadron) and {\tt XPQ(KFL,2)} distributions for
the negative helicity.
The parton distributions are selected by switches described earlier (see Table
\ref{parameters}). The polarised distributions $q_\pm$, $g_\pm$ are
constructed from the
unpolarised ones $q$, $g$ (selected by old {\sc Pythia} switches) and
polarised parts
$\Delta q$ and $\Delta g$ selected by {\tt MSTP(177)} (see Table
\ref{parameters}). Four subroutines have been added to calculate
the polarised distributions. These are
{\tt ALTSTI}\footnote{This subroutine has been written by G.\
Altarelli and J.\ Stirling.
Used with permision from the authors.},
{\tt ROSROA}, {\tt ROSROD}, and {\tt GEHSTI}\footnote{
The three parametrisation of Gehrmann\&Stirling contained in {\tt
GEHSTI} are brandnew and not thouroughly tested yet.}.
The subroutines require data
files {\tt altsti.dat}, {\tt rosroa.dat}, and {\tt partons.dat}. These
files must
be visible to {\tt FORTRAN} {\tt open} statement and therefore
they have to be placed in appropriate directory.
The files are supplied with the program.
The parton distribution for fixed helicity are reconstructed from the
polarised and unpolarised distributions according to
$q_\pm=\frac{1}{2}\left(q\pm\Delta q\right)$.
When the polarised and unpolarised parts are combined together
the program performs the unitarity check -- if the resulting total
distribution
becomes negative for one helicity it is put to zero and the
corresponding result for the other helicity is set to the value of the
unpolarised part.
Only polarised parametrisations for protons are implemented. Neutron
parametrisations are obtained from them by isospin symmetry, the
parametrisations for hyperons are constructed by naive SU(3). Charge
conjugations is used to describe the corresponding antiparticles.

\paragraph{\tt SUBROUTINE
ALTSTI(X,Q2,UPV,DNV,SEA,STR,CHM,BOT,TOP,GLU)}\hfill\break
\noindent{\bf Purpose:} to return $x$ times the polarised parton
distributions evaluated at given $x$ ({\tt X}) and $Q^2$ ({\tt Q2})
according to the parametrisation of Altarelli\&Stir\-ling \cite{altsti}.
{\tt UPV} denotes the valence distribution of up-quarks
$x\Delta u^{\rm val}(x,Q^2)$, {\tt DNV} for down-quarks
$x\Delta d^{\rm val}(x,Q^2)$. {\tt SEA} signifies the sea distribution
$x\Delta q^{\rm sea}(x,Q^2)$. {\tt STR}, {\tt CHM}, {\tt BOT}, and
{\tt TOP} label the distributions for the strange-, charm-, bottom-,
and top-quark $x\Delta q(x,Q^2)$, $q = s, c, b, t$ respectively. Finally
{\tt GLU} marks the gluon distribution $x\Delta g(x,Q^2)$.
\hfill\break
{\bf Remarks:}
{\tt ALTSTI} requires the data file {\tt altsti.dat} which has to be
placed in an appropriate directory. {\tt altsti.dat} is supplied with
this program.

\paragraph{\tt SUBROUTINE ROSROD(X,Q2,XPDF)}\hfill\break
\noindent{\bf Purpose:} to return $x$ times the polarised parton
distributions evaluated at given $x$ ({\tt X}) and $Q^2$ ({\tt Q2})
according to the parametrisation of Ross\&Roberts set d \cite{rosrob}.
{\tt XPDF(-6:6)} contains $x\Delta q(x,Q^2)$ for $q = \bar{t}, \bar{b},
\bar{c}, \bar{s}, \bar{u}, \bar{d}, g, d, u,
s, c, b, t$ in this order.

\paragraph{\tt SUBROUTINE ROSROA(X,Q2,XPDF)}\hfill\break
\noindent{\bf Purpose:} to return $x$ times the polarised parton
distributions evaluated at given $x$ ({\tt X}) and $Q^2$ ({\tt Q2})
according to the parametrisation of Ross\&Roberts set a \cite{rosrob}.
The contents of {\tt XPDF(-6:6)} is explained above.
\hfill\break
{\bf Remarks:}
{\tt ROSROA} requires the data file {\tt rosroa.dat} which has to be
placed in an appropriate directory. {\tt rosroa.dat} is supplied with
this program.

\paragraph{\tt SUBROUTINE GEHSTI(IGFLAG,X,Q2,XDDPR)}\hfill\break
\noindent{\bf Purpose:} to return $x$ times the polarised parton
distributions evaluated at given $x$ ({\tt X}) and $Q^2$ ({\tt Q2})
according to the parametrisation of Gehrmann\&Stir\-ling set a ({\tt
IGFLAG=0}), set b ({\tt IGFLAG=1}) or set c ({\tt IGFLAG=2}).
\cite{gehrmann}.
{\tt XDDPR(-6:6)} contains $x\Delta q(x,Q^2)$ for $q = \bar{t}, \bar{b},
\bar{c}, \bar{s}, \bar{u}, \bar{d}, g, d, u,
s, c, b, t$ in this order.
\hfill\break
{\bf Remarks:} {\tt GEHSTI} requires the program package {\tt parton.f},
consisting of the subroutines {\tt polpar}, {\tt parini}, {\tt
getpar}, {\tt q2low}, {\tt q2high}, {\tt xlow}, and {\tt xhigh}
and the data file {\tt partons.dat}
written by Gehrmann and Stirling and used here with
permission of the authors. For further informations about this package
see \cite{gehrmann}. {\tt partons.dat} has to be
placed in an appropriate directory and is supplied with this program.
The package {\tt parton.f} is contained in this program.

\paragraph{\tt SUBROUTINE DPLIST(MLIST)}\hfill\break
\noindent{\bf Purpose:} to display the polarisations of the particle
in the event listing.\hfill\break
{\bf New parameters:} {\tt KD(I)}
\hfill\break
{\bf Remarks:} {\tt DPLIST} is a modification of the {\sc Jetset}
subroutine {\tt LULIST}.
It is changed to display the
polarisation in the final listing. The sign displayed just
behind the particle code denotes polarisation with respect to
the z-axis. When the sign is
missing the particle has been treated as unpolarised.
The information is taken from the vector {\tt KD(I)} and transformed
accordingly to (`$0$',`$1$',`$2$')$\rightarrow$(`$\ $',`$+$',`$-$').
The following format is chosen (polarisation for
the colliding hadrons, helicity for the partons resp.):
\begin{itemize}
\item = `  ': no polarisation/helicity\\
\item = `$+$': positive polarisation/helicity\\
\item = `$-$': negative polarization/helicity
\end{itemize}

\paragraph{\tt SUBROUTINE DPEDIT(MEDIT)}\hfill\break
\noindent{\bf Purpose:} to compress the vector {\tt KD(I)}, containing
the polarisation information,
properly.\hfill\break
{\bf New parameters:} {\tt KD(I)}
\hfill\break
{\bf Remarks:} {\tt DPEDIT} is a modification of the {\sc Jetset}
subroutine {\tt LUEDIT}.

\section{Examples\label{example}}
In the following we give two examples of a main program for a
simulation with {\sc Sphinx} and show the corresponding results. We
considered longitudinal polarised proton-proton scattering in the CMS at
$\sqrt{s}=200\ {\rm GeV}$ and selected the process $qg\rightarrow qg$
for the partonic interaction.
In the first example both beam and target are polarised in
$+z$-direction, in the second example the target spin is reversed.
With regard to the event listings is has to be mentioned that the
displayed format differs from the real because we removed a
few columns such that it fits in this text. In addition the event
listing is cut after the hard interaction (denoted by $\cdots$), i.\
e.\ behind line 8. In
the omitted part there is no polarisation information and it has the
same format as the original {\sc Pythia} listing.

The information about the polarisation flow is displayed as explained
above right behind the flavour code {\tt KF}. The first row contains the
information about the beam particle. In addition to the {\sc Pythia}
labels the sign `$+$' behind the flavour code for the proton {\tt
KF=2212} denotes the polarisation in positive $z$-direction.
Accordingly the sign `$+$' (`$-$') in the second row signifies the
polarisation of the target in positive (negative) $z$-direction in the
first (second) example. The third and fourth line represents the
initiators of the initial state shower. In both examples they are
gluons with positive helicities for both the beam and the target side.
During the initial state shower the beam side parton becomes an
$\bar{s}$ with positive helicity, whereas the target side parton
remains a positive helicity gluon. These partons undergo the hard
interaction the resulting partons of which are displayed in the lines seven
and eight. In the following the final state interaction takes place,
i.\ e.\ the outgoing partons fragment, the unstable produced hadrons decay,
etc.\, as long as only stable particles exist. In the final state
polarisation is not traced and consequently there is no polarisation
informations about these lines provided. This part of the listing is
then again the same as the corresponding {\sc Pythia} listing.

\subsection{The Main Programs}
\subsubsection{First Example -- parallel polarisation}
\begin{verbatim}
C___    Example of a Main Program for event generating
C___    in longitudinal polarised proton-proton-scattering
C___
C___    E.g.: polarised p(+)p(+)-scattering in CMS
C___          at sqrt(s)=200 GeV
C___
C___    This program has to be linked with
C___    the programs SPHINX and JETSET7.4,
C___    the data files ALTSTI.DAT and ROSROA.DAT,
C___    and the CERN-Libraries.

C___    COMMON BLOCKS of SPHINX for event generation

        COMMON/LUDAT1/MSTU(200),PARU(200),MSTJ(200),PARJ(200)
        COMMON/LUJETS/N,K(4000,5),P(4000,5),V(4000,5)
        COMMON/PYSUBS/MSEL,NSEL,MSUB(200),NSUB(200),
     &  KFIN(2,-40:40),CKIN(200)
        COMMON/PYPARS/MSTP(200),PARP(200),MSTI(200),PARI(200)
        COMMON/PYINT5/NGEN(0:200,3),XSEC(0:200,3)

C==============================================================
C       Polarisation Set-Up
C==============================================================

C___    Polarised simulation
        MSTP(180)=1

C___    Beam positive polarised
        MSTP(171)=1

C___    Target positive polarised
        MSTP(172)=1

C___    Polarised parton distributions a la Altarelli&Stirling
        MSTP(177)=2

C___    Polarised Initial State Shower
        MSTP(176)=1
C___    with polarised parton distributions
        MSTP(175)=1

C==============================================================

C==============================================================
C       Event Set-Up
C==============================================================

C___    Choice of parton processes ``a la carte''
        MSEL=0

C___    Choice of polarised parton processes ``a la carte''
        NSEL=0

C___    Choice of process: qg --> qg
        MSUB(28)=1

C___    Process 28 polarised
        NSUB(28)=1

C___    Number of generated events
        NEVENT=1000

C___    kinematical cuts
C___    P_T-cut (minimum)
        CKIN(3)=3.
C___    P_T-cut (maximum)
        CKIN(4)=25.

C==============================================================

C==============================================================
C       Start of event generation
C==============================================================

C___    Initialisation

        CALL PYINIT('CMS','p','p',200.)

C___    Loop over events
        DO 100 I=1,NEVENT

C___    Event generation
          CALL PYEVNT

C___    List the first event with spininformation
          IF(I.EQ.1) CALL DPLIST(1)

  100     CONTINUE

C___    Print cross section and histogram
        CALL PYSTAT(1)

        END
\end{verbatim}
\subsubsection{Second Example -- antiparallel polarisation}
The second example is constructed by replacing
{\tt MSTP(172)=1} by {\tt MSTP(172)=2} in the first example, i.\ e.\
by switching the spin of the target.
\subsection{The Event Listings}
\subsubsection{First example -- parallel polarisation}
\begin{verbatim}
                          SPHINX
            **  Last date of change:   6 Aug 1994  **

            The Lund Monte Carlo - PYTHIA version 5.7
            **  Last date of change:   3 Apr 1992  **

            The Lund Monte Carlo - JETSET version 7.4
            **  Last date of change:  10 Mar 1992  **

1********** PYINIT: initialization of PYTHIA routines **********

 ===============================================================
 I                                                             I
 I      PYTHIA will be initialized for a p on p collider       I
 I          at    200.000 GeV center-of-mass energy            I
 I                                                             I
 ===============================================================

 *PYMAXI: summary of differential cross-section maximum search *

   ==========================================================
   I                                      I                 I
   I  ISUB  Subprocess name               I  Maximum value  I
   I                                      I                 I
   ==========================================================
   I                                      I                 I
   I   28   f + g -> f + g                I    3.9475E+00   I
   I   96   Semihard QCD 2 -> 2           I    1.7380E+02   I
   I                                      I                 I
   ==========================================================

 ************** PYINIT: initialization completed ***************



                    Event listing (summary)

    I  particle/jet KF orig   p_x     p_y     p_z      E      m

    1  !p+!       2212+   0   0.000   0.000  99.996 100.000  0.938
    2  !p+!       2212+   0   0.000   0.000 -99.996 100.000  0.938
 =================================================================
    3  !g!          21+   1   0.130  -0.241  33.098  33.099  0.000
    4  !g!          21+   2   0.151  -0.503  -3.866   3.902  0.000
    5  !s~!         -3+   3  -0.838  -0.101  27.030  27.043  0.000
    6  !g!          21+   4  -0.036   2.481  -0.350   2.505  0.000
    7  !s~!         -3    0   2.361   2.098  26.049  26.240  0.199
    8  !g!          21    0  -3.235   0.281   0.631   3.308  0.000
 =================================================================
    ...

1** PYSTAT:  Statistics on Number of Events and Cross-sections ***

 =================================================================
 I                              I                    I           I
 I       Subprocess             I  Number of points  I   Sigma   I
 I                              I                    I           I
 I------------------------------I--------------------I    (mb)   I
 I                              I                    I           I
 I N:o Type                     I Generated    Tried I           I
 I                              I                    I           I
 =================================================================
 I                              I                    I           I
 I  0 All included subprocesses I      1000     5855 I 6.695E-01 I
 I 28 f + g -> f + g            I      1000     5855 I 6.695E-01 I
 I                              I                    I           I
 =================================================================

 ** Fraction of events that fail fragmentation cuts =  0.00000 ***
\end{verbatim}
\subsubsection{Second example -- antiparallel polarisation}
\begin{verbatim}
                          SPHINX
            **  Last date of change:   6 Aug 1994  **

            The Lund Monte Carlo - PYTHIA version 5.7
            **  Last date of change:   3 Apr 1992  **

            The Lund Monte Carlo - JETSET version 7.4
            **  Last date of change:  10 Mar 1992  **

1********** PYINIT: initialization of PYTHIA routines **********

 ===============================================================
 I                                                             I
 I      PYTHIA will be initialized for a p on p collider       I
 I          at    200.000 GeV center-of-mass energy            I
 I                                                             I
 ===============================================================

 *PYMAXI: summary of differential cross-section maximum search *

   ==========================================================
   I                                      I                 I
   I  ISUB  Subprocess name               I  Maximum value  I
   I                                      I                 I
   ==========================================================
   I                                      I                 I
   I   28   f + g -> f + g                I    3.9772E+00   I
   I   96   Semihard QCD 2 -> 2           I    1.7380E+02   I
   I                                      I                 I
   ==========================================================

 ************** PYINIT: initialization completed ***************



                    Event listing (summary)

    I  particle/jet KF orig   p_x     p_y     p_z      E      m

    1  !p+!       2212+   0   0.000   0.000  99.996 100.000  0.938
    2  !p+!       2212-   0   0.000   0.000 -99.996 100.000  0.938
 =================================================================
    3  !g!          21+   1   0.784   0.462  33.089  33.102  0.000
    4  !g!          21+   2  -0.502   0.119 -33.887  33.891  0.000
    5  !s~!         -3+   3  -0.240   0.737  21.851  21.865  0.000
    6  !g!          21+   4  -5.178  -4.192   4.698   8.152  0.000
    7  !s~!         -3    0  -4.508   0.563  25.310  25.715  0.199
    8  !g!          21    0  -0.910  -4.018   1.239   4.302  0.000
 =================================================================
    ...

1** PYSTAT:  Statistics on Number of Events and Cross-sections ***

 =================================================================
 I                              I                    I           I
 I       Subprocess             I  Number of points  I   Sigma   I
 I                              I                    I           I
 I------------------------------I--------------------I    (mb)   I
 I                              I                    I           I
 I N:o Type                     I Generated    Tried I           I
 I                              I                    I           I
 =================================================================
 I                              I                    I           I
 I  0 All included subprocesses I      1000     5870 I 6.868E-01 I
 I 28 f + g -> f + g            I      1000     5870 I 6.868E-01 I
 I                              I                    I           I
 =================================================================

 ** Fraction of events that fail fragmentation cuts =  0.00000 ***
\end{verbatim}
\section*{Acknowledgements}
We all thank T.\ Sj\"ostrand very much for his help and encouragement
in understanding the details of the {\sc Pythia} code.
A.\ S.\ thanks the MPI f\"ur Kernphysik for its support.
The work of
L.\ M.\  was supported in part by KBN under grant 2~P302~143~06. This work
was supported also by DFG (Scha 458/3-1) and BMFT-KBN (Projekt X081.92)


\begin{thebibliography}{99}
\bibitem{pyman} T.~Sjostrand, Computer Phys. Commun. 82 (1994) 74
\bibitem{rhic} {\em Proc.\ Seventh Int.\ Conf.\ in
Ultra-Relativistic Nucleus-Nucleus Collisions}, Eds.\ G.\ Baym {\em et al.},
Nucl.\ Phys.\ {\bf A498} (1989);
RHIC Spin Collaboration, {\em Proposal on Spin Physics
Using the RHIC Polarized Collider} August 1992;\\
RHIC Spin Collaboration, UPDATE, Oktober 1993
\bibitem{altsti} G.~Altarelli and J.~Stirling, Particle World 1 (1989) 40.
\bibitem{rosrob} G.\ G.\ Ross and R.\ G.\ Roberts, Rutherford preprint
RAL-90-062 (1990).
\bibitem{altpar} G.\ Altarelli and G.\ Parisi, Nucl.\ Phys.\  {\bf B175}
(1977) 298
\bibitem{gehrmann}T.\ Gehrmann and W. J.\ Stirling, Durham preprint
DTP/94/38 (1994).
\bibitem{crosssec}
R.\ Gastmans and Tai Tsun Wu, {\sl The Ubiquitous Photon --
Helicity Method for QED and QCD}, Oxford Science Publications (1990);
K.\ Hidaka Nucl.\ Phys.\  {\bf B192} (1981) 369
\bibitem{jetset} H.-U.~Bengtsson and T.~Sj\"ostrand, Computer Phys.
Commun. 46 (1987) 43.
\end{thebibliography}
\end{document}